\documentclass[5p,authoryear,times]{elsarticle}

\usepackage{graphicx}
\usepackage{amsmath,amssymb,amsbsy}

\newcommand{\msun}{M_{\odot}}
\newcommand{\mbh}{M_{\bullet}}

\newcommand{\msigma}{\mbh-\s}

\def\barr{\begin{array}}
\def\earr{\end{array}}
\def\berr{\begin{eqnarray}}
\def\err{\end{eqnarray}}
\def\berrno{\begin{eqnarray*}}
\def\errno{\end{eqnarray*}}
\def\be{\begin{equation}}
\def\ee{\end{equation}}
\def\fr{\frac}
\def\la{\langle}
\def\ra{\rangle}

\newcommand{\pol}[1]{\stackrel{\rm LCP}{\mathrm{RCP}}}

\newcommand{\me}{\mathrm{e}}

\def\apjs{{\it ApJS} }
\def\apj{{\it ApJ} }
\def\aj{{\it AJ} }
\def\apjl{{\it ApJL} }

\def\mnras{{\it MNRAS} }
\def\nat{{\it Nature} }

\renewcommand{\o}{\omega}
\renewcommand{\O}{\Omega}

\newcommand{\s}{\sigma}

\renewcommand{\t}{\theta}

\journal{New Astronomy} 
\begin{document} \begin{frontmatter}

\author{M. Safonova} \ead{rita@iiap.res.in} \author{C. S. Stalin} 
\ead{stalin@iiap.res.in} \address{Indian Institute of Astrophysics, Bangalore 560034, India} \cortext[cor1]{Corresponding author. Tel.: +91(80) 2553 0672, 
Fax: +91(80) 2553 4043}

\title{Detection of IMBHs from microlensing in globular clusters}

\begin{abstract}
Globular clusters have been alternatively predicted to host intermediate-mass 
black holes (IMBHs) or nearly impossible to form and retain them in their centres. 
Over the last decade enough theoretical and observational evidence have accumulated to believe
that many galactic globular clusters may host IMBHs in their centres, just like galaxies do. 
The well-established correlations between the supermassive black holes and 
their host galaxies do 
suggest that, in extrapolation, globular clusters (GCs) follow the same relations. Most of 
the attempts in search of the central black holes (BHs) are not direct and present enormous 
observational difficulties due to the crowding of stars in the GC cores. Here we propose a new
method of detection of the central BH --- the microlensing of the cluster stars by the central
BH. If the core of the cluster is resolved, the direct determination of the lensing curve 
and lensing system parameters are possible; if unresolved, the differential imaging technique 
 can be applied. We calculate the optical depth to central BH microlensing 
for a selected list of Galactic GCs and estimate the average time duration of the events.
We present the observational strategy and discuss the detectability of microlensing events
using a 2-m class telescope. 
\end{abstract}

\begin{keyword}
globular clusters \sep central black hole \sep microlensing 
\PACS 98.20.Gm \sep 97.75.De

\end{keyword}

\end{frontmatter}

\section{Introduction}
\label{}

There has been considerable success in detecting the supermassive
black holes (SMBHs) in the Universe. They are the black holes 
(BH) with masses in the range $10^6-10^{9.5}\,\msun$, 
called so to distinguish them from the stellar mass 
black holes produced by the death of massive stars. There is no 
dearth of stellar-mass ($1-15 \,\msun$) black holes either; by some estimates there may be
$10^7-10^9$ in every galaxy (Shapiro \& Teukolsky 1983; Brown \& Bethe 1994). 
Black holes with masses in the range 
$10^2-10^4\,\msun$, appropriately called the intermediate-mass black holes (IMBHs), 
however, remain a mystery. IMBHs have persistently evaded discovery in spite of 
considerable theoretical and observational efforts. Dense star clusters, such as
globular clusters, have long been suspected as possible 
sites for the formation of IMBHs. The idea
that some, if not all, globular clusters can host a central black hole
actually preceedes the notion of the supermassive black holes 
\citep{Frank-Rees}, and more than thirty years ago attempts were 
made to discover them by their X-ray emission \citep{Bahcall-Ostriker}.  
Hunting for globular-cluster black holes was recognized as a task suited for
{\it HST}'s exquisite resolution which is needed to look close to a black hole.
This idea was restimulated by the capability of {\it Chandra X-ray Observatory} to
resolve sensitively the X-ray emission from the very centres of globular
clusters; and recent detection of X-ray emission in the globular cluster G1 
in M31
\citep{G1xray} provided additional clue to the existence of an 
IMBH in this cluster. The growing evidence that some Galactic globular clusters
could harbour central black holes, just as galaxies do, stimulates the searches 
and development of new methods for proving their existence. 

\subsection{Theoretical evidences of IMBH in GCs}

Portegies \& McMillan (2002) showed that a runaway merger among the most
massive stars in the globular cluster leads to the formation of an IMBH, provided that
the core collapse proceeds faster than their main-sequence lifetime. For a
globular cluster that evolves in the Galactic tidal field, the corresponding present-day
half-mass relaxation time would have to be $10^8$ years \citep{Marel2003}. Many of
the Milky Way globular clusters have half-mass relaxation times in the range 
$10^8-10^9$ years, and some below $10^8$ years \citep{Harris}. 
Another possible way for the formation of IMBHs in globular clusters is through the repeated
merging of compact objects, if, for example, a single $\sim50\,\msun$ 
BH were initially somewhere in the cluster, it would sink to the centre
through dynamical friction, and slowly grow in mass through merging with stellar-mass
black holes \citep{Miller-Hamilton}. Clusters with central densities $\gtrsim 10^5$ pc$^{-3}$ 
will have
high enough encounter rates to produce $10^2-10^4\,\msun$ BHs. In the
Milky Way it would imply that roughly 40\% of globulars could host such objects.

Other, more exotic scenarios were suggested, such as direct collapse of population
III stars and subsequent growth by accretion \citep{Madau-Rees}, or the accretion of
supernovae winds funneled by radiation drag exerted by stars on the interstellar medium,
onto the cluster centre, forming the central massive object which eventually collapses
to form an IMBH \citep{kawakatu-umemura2005}. While the first scenario does not 
require the host stellar system to be dense, the latter rules out the formation
of the central BHs in most of the present-day galactic GCs 
on the basis of their insufficient total mass and/or central velocity dispersion. 

\subsection{Observational evidences for IMBH in GCs}

\begin{figure}
\begin{center}
\includegraphics[width=70mm,height=70mm,angle=-90]{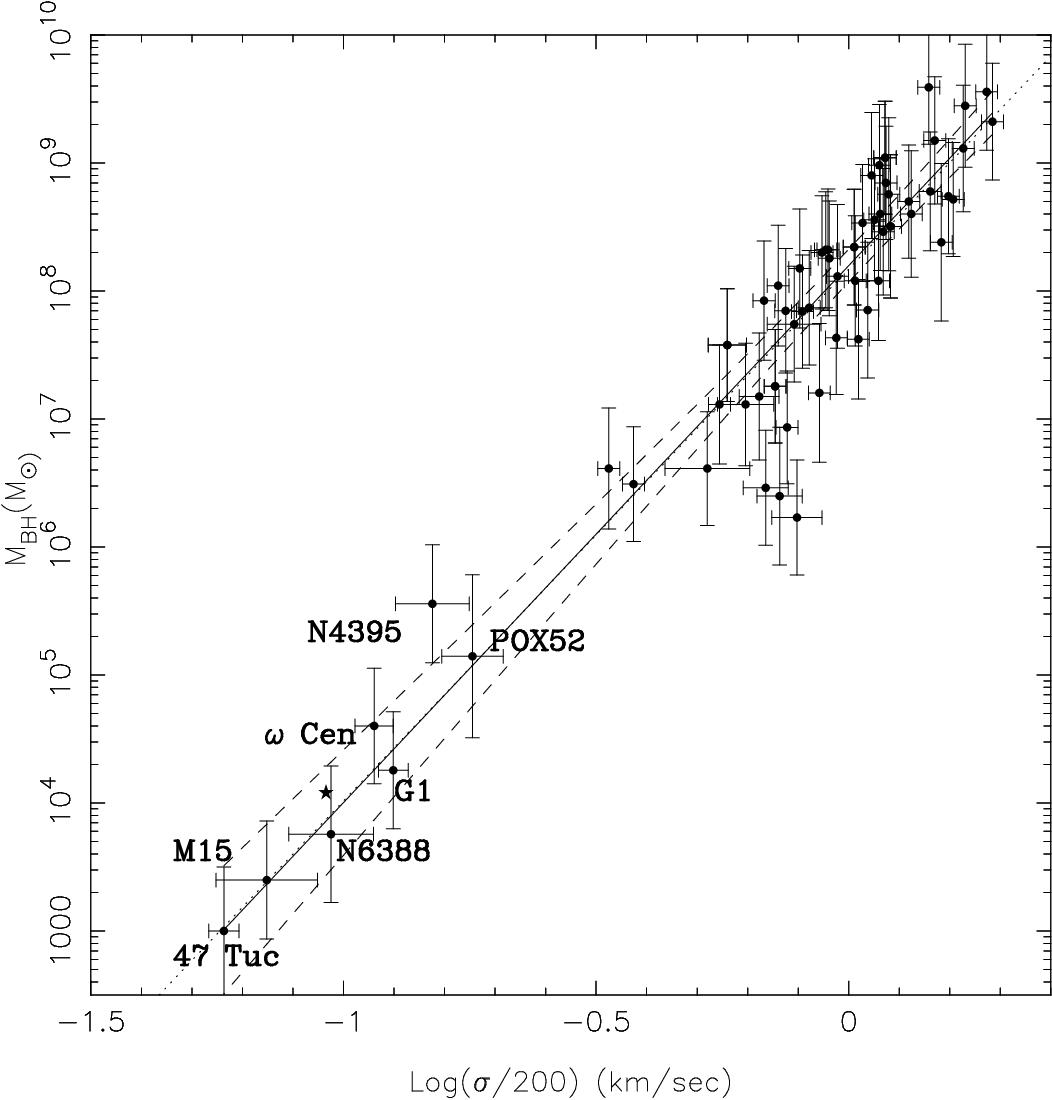}
\caption{\small Mass of the central black hole $\mbh$ vs. central velocity 
dispersion $\s$ for a sample consisting of 49 galaxies from 
G\"{u}ltekin et al. (2009), NGC4395 \citep{4395} and Pox 52, the lowest-mass AGN to date \citep{pox52}, and five globular clusters (47Tuc, $\o$ Cen, NGC6388, G1 and M15 \citep{{47Tuc},{oCen},{6388}, {G1},{Bosch2006}}). 
Dotted line is the linear regression fit for only galaxies and solid line 
is the fit including the globular clusters. Dashed line is the 2-sigma error on the 
regression. Single star is the most recent estimate of the $\omega$ Cen black 
hole mass \citep{Anderson2009}.} 
\label{fig:correlations}
\end{center}
\end{figure}

Several GCs were suggested to harbour IMBHs 
at their centres. The main evidence  
comes from the analysis of the central velocity dispersions of some GCs
\citep{{Gerssen2002},{Gerssen2003},{G1}}, and, in some cases, from 
the presence of a rotation in the core \citep{{Bosch2006},{Gebhardt2000}}.
Suggested upper limits of radio emission from low luminosity IMBHs 
in GCs (Maccarone et al. 2005) have led to upper estimates of the central BH mass 
in two nearby galactic GCs, 
47 Tuc and NGC 6397. Detection of radio (Ulvestad et al. 2007) and
X-ray (Pooley \& Rappaport 2006) from the GC G1 in M31 also supports
the previous claims of Gebhardt et al. (2005) that it hosts a central BH.
These IMBHs lie on the extrapolated $\msigma$ relation found for SMBHs in galactic nuclei 
(see Fig.~\ref{fig:correlations}), and this leads to a prediction of a 
central mass of  $\sim 10^3\,\msun$ for a typical globular cluster
having a velocity dispersion of the order of 10 km/sec.\citep{{Magorrian98},{Kormendy-Richstone},{Marel1999}}. 
  
Despite the observational evidence of IMBH in GCs, 
there remains considerable debate about whether the reported excess
of mass in the centres of GCs can be well modelled by
a cluster of low-mass objects, such as white dwarfs (WD), neutron stars (NS)
or stellar-mass BHs, rather than an IMBH \citep{Baumgardt2003a}. 
None of the existing methods can distinguish between these 
two alternatives, mainly because the fitting procedure is relatively insensitive to 
the precise nature of the dark matter contained within the innermost 
region of the cluster \citep{Baumgardt2003b}. Detection of gravitational 
waves from an IMBH can be a potential method for resolving this 
argument, but for that the central IMBH has to be a binary and at final 
stages of coalescence. 
{\it Gravitational microlensing} of a background star by the central black hole,
on the other hand, can in principle be a clear 
diagnostic tool as there is a significant difference in the lensing signatures of 
a single point-mass lens (a single IMBH), a binary lens, and, most importantly, of 
an ensemble of point-mass lenses (which would be the case if the central 
mass consisted of a conglomeration of low-mass objects). Thus the detection of a 
lensed signal in the centre of a GC can resolve the controversy about the 
nature of a central dark mass detected in some globulars and prove beyond
doubt the existence of IMBH in GCs.

Paczynski (1994) was the first to 
suggest that lensing of stars in SMC or Galactic bulge can reveal 
compact objects in foreground GCs. However it was found that the probability of such 
events is low and moreover such clusters are too few. It was also suggested to 
use GC stars as sources for halo lenses
to distinguish between different galactic halo models (Gyuk \& Holder 1998; 
Rhoads \& Malhotra 1998). This would require monitoring each stars in a globular
cluster or on average over all globular clusters about 60,000 stars. Both suggested
methods would need very powerful telescopes with high resolution imaging.

In this paper we propose to consider the microlensing events 
that are expected when globular cluster stars pass behind the central BH 
that acts as a lens, inducing amplification of light. We show here that 
by observing enough number of globular clusters there is a 
chance to prove the existence of an IMBH. This method removes some 
of the ambiguities usually present in the galactic microlensing events, 
because in GCs the location of both the lens and the source, and their
velocities are well constrained.

\section{Globular Clusters Microlensing}
 
\subsection{Microlensing: Overview}

According to standard microlensing theory, when a background source 
is sufficiently close to the line of sight to a lens, its apparent 
brightness is increased by the factor 
\be 
A=\fr{u^2+2}{u(u^2+4)^{1/2}}\,,\quad
u^2(t)=u^2_{\rm min}+\left(\fr{t-t_{\rm max}} {t_{\rm
E}}\right)^2\,,
\ee
where $u_{\rm min}$ is the impact parameter in terms of the Einstein
radius, $t_{\rm max}$ is the time at which maximum amplification occurs, 
and $t_{\rm E}$ is the time it takes the source to move
across the Einstein radius, which is given by \be R_{\rm
E}=\left[\fr{4G\mbh}{c^2}\fr{D_{\rm L}D_{\rm LS}}{D_{\rm S}}
\right]^{1/2}\,,
\ee
where $D_{\rm S}$, $D_{\rm L}$ and $D_{\rm LS}$ are the 
observer-source, observer-lens and lens-source distances, respectively,
and $\mbh$ is the mass of the lens (black hole). {\bf The resulting light 
curve is symmetric around $t_{\rm max}$ and achromatic (the light curve
shape is independent of wavelength)}. We take the duration 
of the microlensing event
$t_{\rm E}$ as the time it takes a source to cross the Einstein
radius, 
\be 
t_{\rm E}=\fr{R_{\rm E}}{v_{\bot}}\,,
\label{eq:duration}
\ee
where ${v}_{\bot}$ is the relative two-dimenstional transverse speed, 
$v^2_{\bot}=v^2_x+v^2_y$, where $v_x$ and $v_y$ are the components 
of the source velocity perpendicular to the line of sight. The optical depth of 
the microlensing is the probability that a background source lies 
inside the Einstein radius of the lens,
\be \tau=\int \pi R_{\rm E}^2\,n(r) dr\,,
\label{eq:tau}
\ee
where $n(r)$ is the number density of the sources.

We distinguish two types of microlensing in globular clusters: lensing of GC stars 
by the central IMBH and self-lensing of GC stars by the stars in the same cluster.
 
\subsection{Microlensing due to the central IMBH}
\label{sec:M15}

In the case when both the source and the lens are situated in the globular
cluster we have $D_{\rm LS} \ll D_{\rm L} \approx D_{\rm S}$, and the
Einstein radius is 
\be
R_{\rm E}\approx (2.8\, {\rm
AU})\left(\fr{\mbh}{\msun}\right)^{1/2} \left(\fr{D_{\rm
LS}}{1\, {\rm Kpc}}\right)^{1/2}\,.
\ee
The calculation of the optical depth in 
this case is a little bit different from the conventional one. Here
we have a lens whose Einstein radius is a function of the distance $r$ between
the lens and a background source star, $R_{\rm E}(r)$. For an observer
far away from the lens we consider a {\bf cone} with a cross-section of 
{\bf $\pi R_{\rm E}(r)^2$} and a length $D_{\rm LS}$. {\bf The probability
that at any given instant there is a star inside the lensing cross-section is}
\be dN_{\rm event}=\fr{\rho(r)}{M_{*}}\pi R_{\rm E}^2(r) dr, 
\label{number}
\ee
where $\rho(r)$ is the density of GC and $M_{*}$ is the average mass 
of stars, which we assume to be equal to the solar mass. Integrating over $r$
{\bf gives the number of ongoing microlensing events, $N_{\rm event}$ at 
any given instant}. 
Unlike the standard optical depth (Eq.~\ref{eq:tau}), this optical depth
depends on the sources mass function.

{\bf For the mass distribution in
globular clusters we use the Plummer density profile} \citep{Binney-Tremaine},
\be \rho(r)=\rho_0
\fr{1}{\left[1+\left(r/r_c\right)^2\right]^{5/2}}\,,
\label{cgmodel}
\ee 
where $\rho_0$ is the central mass density and
$r_c$ the core radius.

As an illustration we give here the details of calculations for M15, a globular cluster 
for which a central black hole mass of $2.3\times 10^3\,\msun$ was 
given as the most probable value \citep{Kiselev}.  
For M15, $\rho_o\simeq 7.4\times 10^6\,\msun$ pc$^{-3}$ \citep{Bosch2006} and 
$r_c\simeq 0.2$ pc \cite{DePaolis}, Eq.~\ref{number}
gives the number of microlensing events as $N_{\rm event} =1.43\times 10^{-4}$. 
This would mean that monitoring the centres of about
$10^3$ globular clusters {\bf theoretically} gives a ten percent chance of 
seeing 
a microlensing event in progress. 

In order to estimate the number of events during a given
observational time $T_{\rm obs}$, we again consider a {\bf tube} with 
diameter $2R_{\rm E}(r)$ and length $v_{\rm S} T_{\rm obs}$, 
where $v_{\rm S}$ is the relative transverse velocity of a star
residing at a distance $D_{\rm S}$. {\bf This is given as} 
\begin{equation}
N_{\rm obs} = \int_{0}^{\infty} 2\pi R_{\rm E}(r) v_{\rm S} T_{\rm obs}\, n(r) dr\,.
\end{equation}
Using Eq.~\ref{cgmodel} for the distribution of the matter 
in GC, the number of microlensing events within one year of observations
is $N_{\rm obs}=10^{-3}$. We may expect that by observing $\sim 100$ clusters similar
to M15 for ten years, 
one can detect one microlensing event due to the IMBH.

We assume that the two components of $\mathbf{v}_{\bot}$ are normally
disrtributed. The probability that the velocity is in the range  
$\mathbf{v}_{\bot}$ and $\mathbf{v}_{\bot} + d\mathbf{v}_{\bot}$ is given by 
\be
dp(v_x,v_y) = \frac{1}{2\pi\s^2} \exp{\left( -\frac{v^2_x+v^2_y}{2\sigma^2}\right)} 
dv_x\,dv_y\,.
\label{eq:prob}
\ee
Due to cylindrical symmetry, we integrate this probability in the azimuthal
direction to obtain 
\be
dp(v) = \frac{1}{\s^2} \exp{\left( -\frac{v^2}{2\sigma^2}\right)} dv\,,
\label{eq:prob1}
\ee
where $v=|\mathbf{v}_{\bot}|$. It can be easily verified that the most 
probable speed is $v=\s$. 

To estimate the mean duration of events we divide the mean Einstein radius
\be
\la R_{\rm E}\ra=\fr{\int_0^{\infty}P(r)R_{\rm E}(r)dr}{\int_0^{\infty}P(r)dr} 
\ee 
by the the most probable speed $v=\s$ of stars in the cluster, 
\be
\la t_{\rm E}\ra =\fr{\la R_{\rm E}\ra}{\s}\,,
\label{eq:t_E}
\ee
where $P(r)=dN_{\rm event}/dr$ is the probability of a star to be inside 
the Einstein ring. For example in the case of M15, the mean 
Einstein radius is 2.07 AU, and with the central velocity dispersion 
$\s=14.1$ km/sec, the mean duration of event is $t_{\rm E}=217$ days.
 
Apart from the mean duration of events, we can also obtain the distribution 
of event time-scales. For this purpose let us denote the event time-scale 
corresponding to $v$ as $\la t_{\rm E}\ra\left(v\right)$ and its inverse function
as $v\left(\la t_{\rm E}\ra\right)$. The cumulative probability that the 
event time-scale is less than $\la t_{\rm E}\ra$ is given by
\be
P\left( < \la t_{\rm E}\ra\right)=\fr{1}{\s^2} 
\int_{v(\la t_{\rm E}\ra)}^{\infty} v 
\me^{-v^2/2\s^2}d v\,.
\label{eq:veloprobability}
\ee

\begin{figure}
\begin{center}
\includegraphics[width=8cm,height=5cm,angle=0]{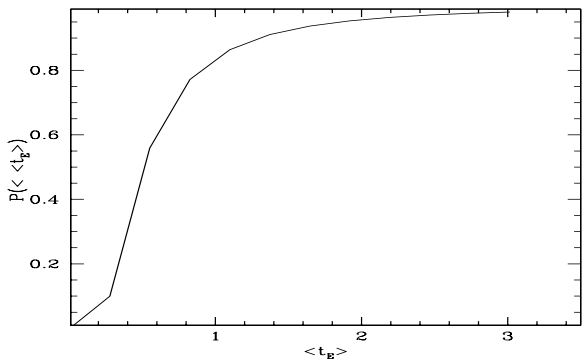}
\caption{Cumulative probability that event time-scale is
less than or equal to $\la t_{\rm E}\ra$.  }
\label{fig:timeprob}
\end{center}
\end{figure}

The Figure~\ref{fig:timeprob} is a plot of $P(< \la t_{\rm E}\ra)$ vs 
$\la t_{\rm E}\ra$ in years for the globular cluster M15. We can see from
this Figure that 90\% of the events last less than
about two years. The fraction of events between $t_1$ and $t_2$ ($t_2>t_1$) 
is given by the difference $P( < t_2)-P(< t_1)$.

\subsection{Self-Lensing}

In case of self-lensing, each star can be both a lens and a source,
and therefore the optical depth shall be integrated over the cluster. We call this optical depth 
an integrated optical depth. Let us take the polar coordinate system with the $z$-axis 
along the optical axis. The number of lenses $dN_{\rm L}$ and sources $dN_{\rm S}$
is
\be
dN_{\rm L} = \frac{\rho(z_{\rm L,r})}{M}r dr d\theta dz_{\rm L}\,,\quad
dN_{\rm S} = \frac{\rho(z_{\rm S,r})}{M}r dr d\theta dz_{\rm S}\,
\ee
respectively. We calculate the number of ongoing microlensing events by multiplying the 
overall Einstein radius covering the observational field by the number of the source stars
\begin{equation}
dN_{\rm self} = \frac{\pi R_{\rm E}^2 d N_{\rm L}}{A}dN_{\rm S},
\end{equation}
where $A$ is the GC projected area. Since $D_{\rm L}\simeq D_{\rm S}$ in the GC
microlensing, the number of the ongoing microlensing events is
\be
\begin{split}
&N_{\rm self} = \frac{4\pi^2 GM_*(z_{\rm L}-z_{\rm S})}{c^2}
\left(\frac{2}{M_*L}\right)^2 \\
& \times \int\limits_{z_{\rm L} = -L}^{z_{\rm L}= L}
\int\limits_{r=0}^{r=L}
\rho(r,z_{\rm L})r dr dz_{\rm L}   
\times \int\limits_{z_{\rm S} = z_{\rm L}}^{z_{\rm S}=L}
\int\limits_{r'=0}^{r'=L}
\rho(r',z_{\rm L})r' dr' dz_{\rm S} \,,
\end{split}
\label{eq:self-lensing}
\ee
where $L$ is the size of the globular cluster.
To have a rough estimate of the number of {\bf self-lensing} events,
we take the uniform distribution of stars in GC, then
Eq.~(\ref{eq:self-lensing}) gives
\begin{equation}
N_{\rm self}  = \frac{R_{\rm S}}{L}
\left(\frac{M_{\rm GC}}{M_\ast}\right)^2,
\end{equation}
where $R_{\rm S}$ is the Schwarzschild radius of a GC star,
$M_{\rm GC}$ is the mass of the GC, and $M_\ast$ is the mass of
the cluster star assumed to be solar.
Taking $M_{\rm GC} \approx 10^6 \msun$ and $L$ of the
order of parsec, we obtain the number of self-lensing events 
$N_{\rm self} \simeq 10^{-3}$. 
In one year of observations, the number of events is
about $5\times 10^{-2}$, and in observing $20$ globular clusters
one would expect to see one event. 
One of the characteristics 
of the self-lensing inside the GC is that the
Einstein radius is about $100$ times smaller than that of IMBH and
so we expect the duration of self-lensing events to be of the order
$10$ days. 

\subsection{Globular Cluster Sample}\label{sec:choice}

It has been widely believed that middle-range black holes 
can only reside in the most centrally concentrated
clusters, with density profile very close to that predicted for the 
core-collapse. 
The presence of a BH induces the formation of a density cusp, 
which was found in about 20\% of galactic globular
clusters \citep{Djorgovski-King}. 
For example, M15, where the detection of
IMBH was tentatively reported, has been long known to be 
a proto-typical core-collapsed cluster \cite{Djorgovski-King,Lugger87}.
However, it was argued (Baumgardt et al. 2005) that, on the contrary, 
no core-collapsed cluster can harbour a central BH, as it 
would quickly puff up the core by enhancing the rate of close 
encounters, and one has to look for the clusters with large core 
radius and just a slight slope of
the density profile in the core region. This was further reinforced 
by Miocchi (2007), who, however, made several assumptions that do not 
reflect realistic cluster models \citep{heggie2007}. On the other hand, 
many of the previously thought 
core-collapsed (or post-core-collapsed; PCC) clusters were recently 
shown to posses cores, and King models also appeared to poorly 
represent most globulars in their cores 
\citep{NoyolaGebhardt2006}. 
We therefore, have included in our sample all proposed
Galactic core-collapsed clusters (or PCC) \cite{Trager95}, as well as 
candidates from Baumgardt et al. (2005) (Baumgardt's set). 
Both sets are emphasized by a bold face
number in the Reference column of the Table~1. It is obvious that this
debate is not settled yet, and 
improving observational techniques and devising new
tests, like the one proposed here, could resolve this
long-standing issue.\footnote{Recently Hurley (2007) issued a cautionary
note on using large $r_{\rm c}$ as an indicator of an IMBH presence since
other factors, such as the presence of a stellar BH-BH binary in the
core, can flatten the measured luminosity profile and enlarge the core radius,
and that it is still too early to abandon the earlier used \citep{G1}
IMBH indicators such as the steepening of a M/L ratio in GC cores.}
We also included eleven massive Galactic globular clusters and the 
most massive globular cluster of M31---G1.  This includes $\o$ Cen, 
and NGC6388 with the possible detection of a $5\times 10^3\,\msun$ IMBH \citep{6388}. 
$\o$ Cen and G1 are the most secured
IMBHs detections to date; with the latest value for $\o$ Cen reduced from 
$4.0\times10^4\,\msun$ \citep{oCen} to $1.2\times 10^4 \,\msun$ \citep{Anderson2009}, and 
a reported value for G1 of  $1.8\times 10^4\,\msun$ \citep{G1}.

\begin{figure}
\begin{center}
\includegraphics[scale=0.3]{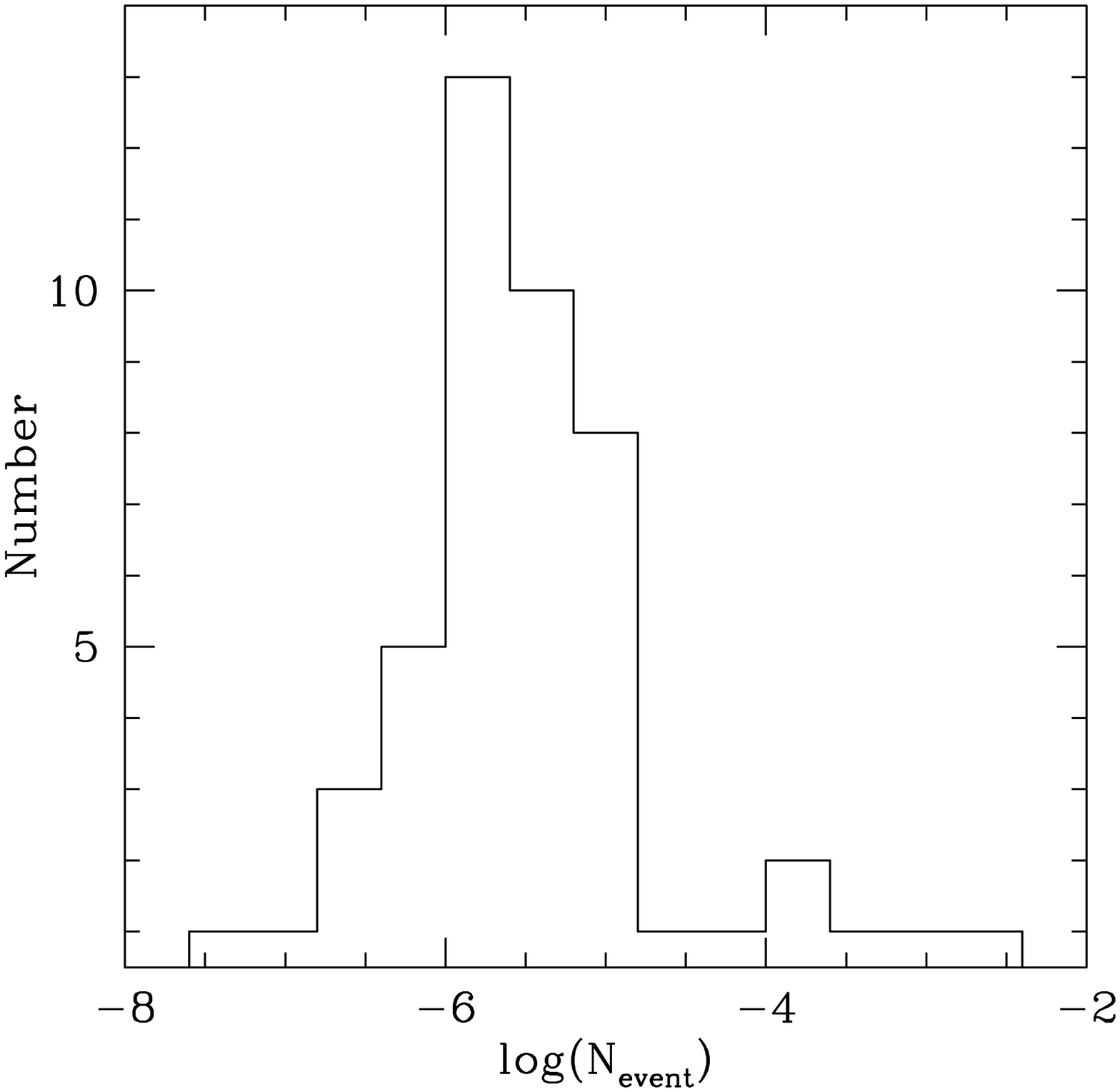}
\includegraphics[scale=0.3]{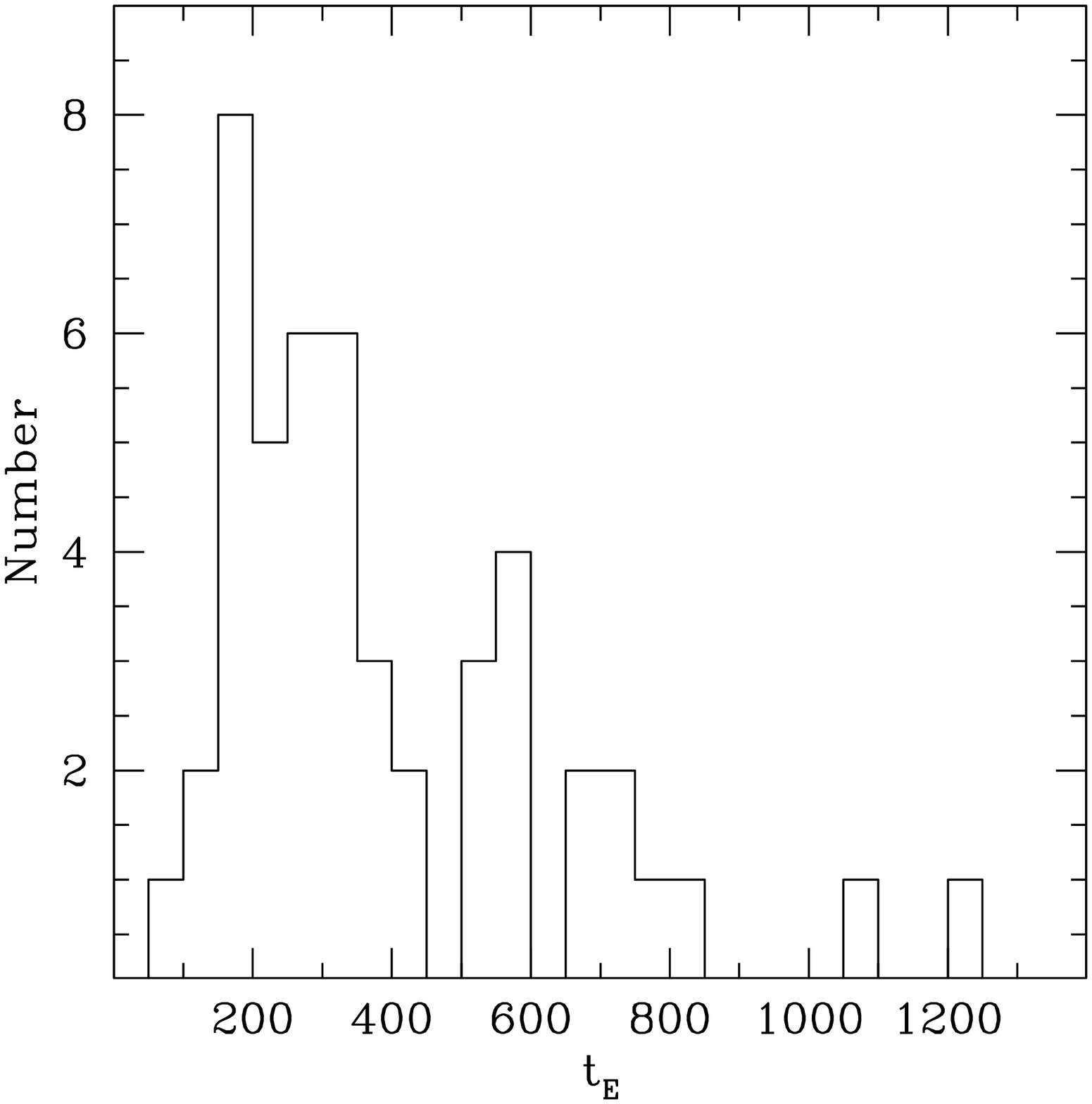}
\caption{Distribution of $N_{\rm event}$ (top panel) and $t_{\rm E}$ 
(bottom panel) for our sample of GCs} 
\label{fig:timeeveprob}
\end{center}
\end{figure}

For our sample of GCs we have calculated the number of events $N_{\rm event}$
and the mean duration of event $\la t_{\rm E} \ra$ (The respective 
histograms are given in Fig.~\ref{fig:timeeveprob}).  
In these 
calculations we assumed the mass of the central black hole of $10^3\,\msun$ for 
all candidates except 
M15 (see Sec.~\ref{sec:M15}), NGC6388, $\o$~Cen and G1 (see above). This choice is 
motivated by  extrapolation of 
the $\msigma$ correlation known for galaxies to globular clusters (see Fig.~\ref{fig:correlations}). 
For $\o$~Cen, instead of the core radius, we assumed the radius of
influence of $\sim 10^4 \,\msun$ BH ($\approx 15^{\prime\prime}$).  
The results are presented in Table~1. 
We also give in Table~1 the heliocentric distance $R_{\odot}$ to 
the cluster, the central velocity dispersion $\s$, the logarithm of the central concentration 
$\rho_0$, the core radii in arcminutes ($\t_c$) and in parsecs ($r_c$). 
The last two columns give the cluster classification type (see Sec.~\ref{sec:classification}) and 
references to the data. 

\section{Discussion}

\subsection{Classification of Globular Clusters.}\label{sec:classification}
Globular clusters are grouped into different classes according to 
their metallicity and horizontal branch morphology (Zinn 1993; Mackey \& van 
den Bergh 2005) namely bulge/disk (BD), old halo (OH), young halo (YH) clusters.
One more class is listed in Table~1, SC, which means that this globular cluster 
is a stripped core of the former spheroidal or elliptical dwarf 
galaxy. It has been 
suggested that $\o$ Cen, M54 and G1 are probably not genuine globular 
clusters, but are the nuclei of accreted galaxies (see \citet{Mackey-Bergh} 
and references therein)
We find that majority of the candidates belong to the OH/BD group and all
SC clusters are also in our Table. 
 Both BD and OH populations
have been strongly modified by tidal forces and bulge and disk shocks and
so these subsystems
consist of mostly compact, higher luminosity and
higher surface brightness clusters, whereas YH group contains a significant 
fraction of extended, diffuse and low-luminosity clusters.

\begin{figure}
\begin{center}
\includegraphics[width=84mm]{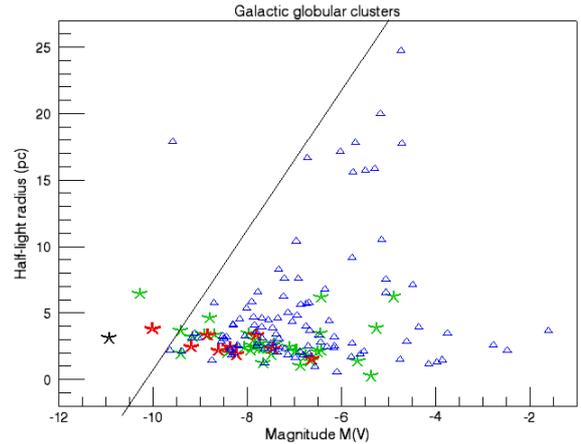}
\caption{$R_{h}$ vs $M_V$ for 146 Galactic globular clusters. The
Baumgardt's set is marked by red asterisks; the CC set by green
asterisks; the remaining clusters by blue triangles. The M31
globular cluster G1 is marked by a black asterisk, the measurement
for its $R_{\rm h}$ is taken from Barmby et al. (2002).
Above the slanted line the red asterisk marks $\o$ Cen, the green
asterisk M54.} 
\label{fig:2}
\end{center}
\end{figure}

In Fig.~\ref{fig:2} we show $R_{\rm h}$, the radius that contains half of 
the cluster stars in projection, versus $M_V$, the integrated luminosity, 
for 146 Galactic GCs (Mackey and van den Bergh 2005). 
Core collapse GCs are marked by green asterisks, the Baumgaurdt's list of GCs
are shown as red asteristicks and the remaining GCs are shown as blue
triangles. There is a sharp edge to the main
distribution of the clusters, called the Shapley line (van den Bergh 2008),
and only 
4 clusters (all belonging to the SC class) lie above the line
(3 galactic GC and G1 from M31). We see that our sample
of clusters is concentrated in a small area of the plot and, as far as
this distribution is concerned, there is no considerable difference 
between the Baumgardt set and CC set. 
It was noticed
\cite{G1} that to the extent that a massive, bound cluster can be
viewed as a `mini-bulge', it may be that every dense stellar system (small or large)
hosts a central black hole.
It is possible that 
there may be some previously unrecognized connection between the formation and 
evolution of globular clusters and their central black holes. 
Loosely bound clusters are indeed susceptible to strong
evolution  \cite{Gnedin-Ostriker}, while compact clusters are
significantly more stable. 

It is not known how significantly the evolutionary effects 
may influence the formation of the cluster's central BH, or the survival of a 
cluster with a central IMBH. In \citep{Baumgardt-Makino} it was proposed on the 
basis of N-body simulations of realistic mutli-mass star clusters
that central IMBH speeds up the dissolution of a star cluster, especially 
if a cluster is surrounded by a tidal field. This would rule out the presence of central
IMBH in BD and OH population, though not in YH. However, one YH cluster with reported
central black hole is a core-collapsed cluster M15  and
core-collapsed clusters are
also ruled out by 
Baumgardt et al. (2004). 
We intend to develop some criteria to select a set of clusters for 
our observational program.
It is tempting to speculate that it is 
worth to look for IMBH in dense, compact and high luminosity, 
and may be old, clusters 
rather than diffuse and low luminosity ones. 
Clearly, there is a contradiction between some 
theoretical approaches (for ex., Baumgardt et al. (2004), Baumgardt et al. (2005)) and 
observational reports, which only stimulates more efforts to resolve the  
issue of the presence of IMBHs in different types of GCs.

\subsection{Selection of Targets and Observational Strategy}

Ideally, we would monitor all Galactic globular clusters.
However, in practice it is a very difficult task.
Even in our selected candidates list (Table 1) it is found
that the optical
depth varies considerably. Besides, the tentative results of
Sec.~\ref{sec:classification} show that it might be useless to look for
diffuse, faint and distant clusters. 
Moreover, in
PCC clusters, if we take into account the mass segregation effect, and/or a mass
distribution law more concentrated towards the centre, a $r^{-7/4}$ profile, the
lensing rate increases. For example, for nearby 47 Tuc, NGC 6397, and NGC 6752
it would nearly double ($18.6\times 10^{-6}$, $7.9\times 10^{-6}$ and $2.6\times
10^{-6}$, respectively).

\begin{figure}
\begin{center}
\includegraphics[scale=0.5]{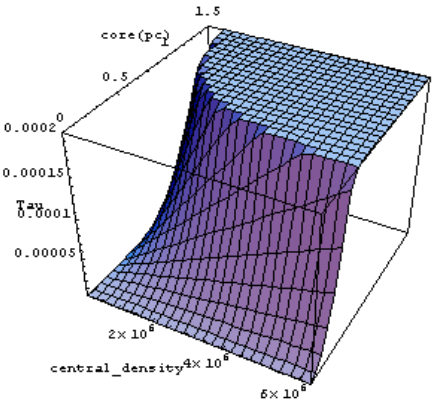}
\vskip -0.5in
\caption{Dependence of optical depth (Tau) on central density and core radius}
\label{fig:5}
\end{center}
\end{figure}

Assuming the threshold optical depth for choosing the observational
targets as $1\times 10^{-6}$, we plot in Figure~\ref{fig:5} the dependence
of the optical depth on core radius and central density. In order to have larger 
optical depth, a cluster has to have either large core radius or/and large 
central density, though the optical depth rises faster with the central density 
rather than the core radius. According to Fig.~4 and section 3.1, 
dense compact clusters are potentially better targets for monitoring programs 
aimed at detecting IMBHs. 

Our observational program suggests monitoring the   
cores (core radius) of selected
globular clusters with the frequency of once a month in two filter-bands
V and I using 2m class telescope. This particular choice of filters is to
eliminate chromatic false events.  
The advantage of this proposed  method is that it offers a direct
relationship between the lens mass and the timescale of the ML event.
In a typical ML event, only $t_{\rm E}$ can be measured
directly and if the distance to the source is known, the remaining
three physical parameters, lens mass $\mbh$, distance to the lens
$D_{\rm L}$ and transverse velocity $v$, are in a degenerate combination,

\be
t_{\rm E} [secs] =  \frac{1.4}{v} \left(\frac{\mbh}{\msun}\right)^{1/2} \left(D_{\rm L}\left(1-\frac{D_{\rm L}}{D_{\rm S}}\right)\right)^{1/2},.
\ee 

For example, in the Galactic ML events $D_{\rm L}$ is not known and 
$v$ can be anywhere
in the range $0 < v < 600$ km/sec, resulting in a distribution of 
lens masses from $0.1 \lesssim \mbh \lesssim 10 \, \msun$ \citep{wamb}.
In a GC microlensing, no star not belonging to a cluster can be a
source, thus even if the source is not directly detected, its tranverse
velocity is within the range of cluster's velocity dispersion (Sect. 2.2). 
The same is true for the distance to the source; distance to the lens
is assumed to the cluster distance. Moreover, lens mass is constrained
theoretically (Sect. 1.2) and source mass is constrained by  
the cluster mass function.
\subsection{Prospects for Detection}

The requirements of a traditional absolute photometry argue that cluster 
ML observations shall be carried with a large telescope with subarcsecond 
seeing \cite{gyukholder}. However, the cores of most GCs are not resolved 
even with an {\it HST} and, besides, it is not necessary to monitor every 
star in a globular cluster; the differential imaging analysis (DIA) takes 
the practice of crowded field photometry to its extreme limit. DIA is 
sensitive to ML events even when the source star is too faint to be detected
at the baseline (Bond et al. 2001). We do not expect to be able to resolve 
the source star in the crowded centres of globular clusters, but with a 
careful application of DIA we do expect to be able to detect the star during any 
suitably bright outburst episode.

To address the question of  detectability of microlensing events in our sample 
of IMBH globular clusters (Table~1), we again take the case of the cluster M15 as an example. 
M15 was observed  for feasibility  in April 2008. The observations were 
carried out on the 2-m Himalayan Chandra Telescope ({\it HCT}) of the Indian Institute of 
Astrophysics, equipped with a 2K$\times$2K CCD, in $V$ and $I$ bands. 

\begin{figure}
\begin{center}
\includegraphics[scale=0.3]{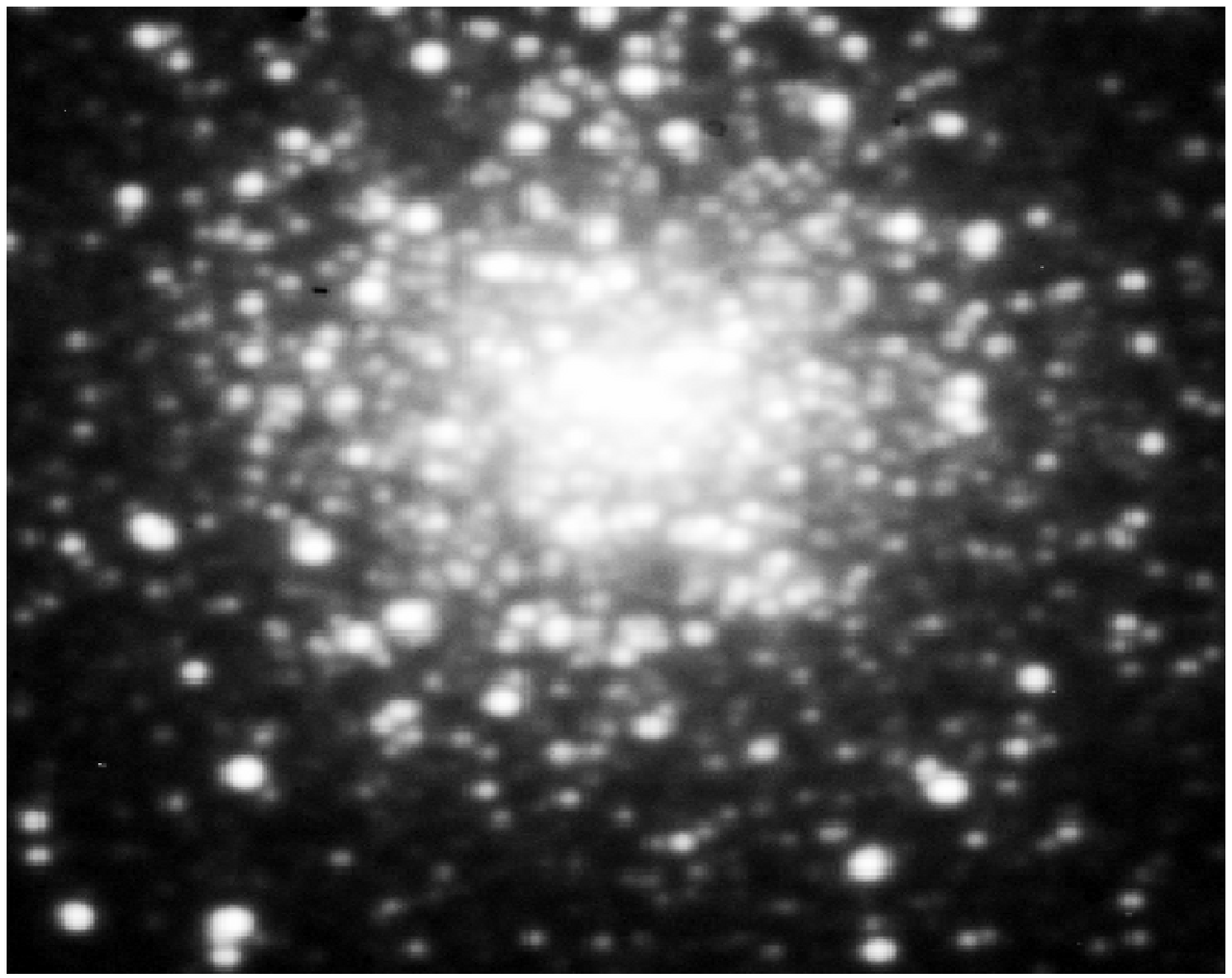}
\includegraphics[scale=0.3]{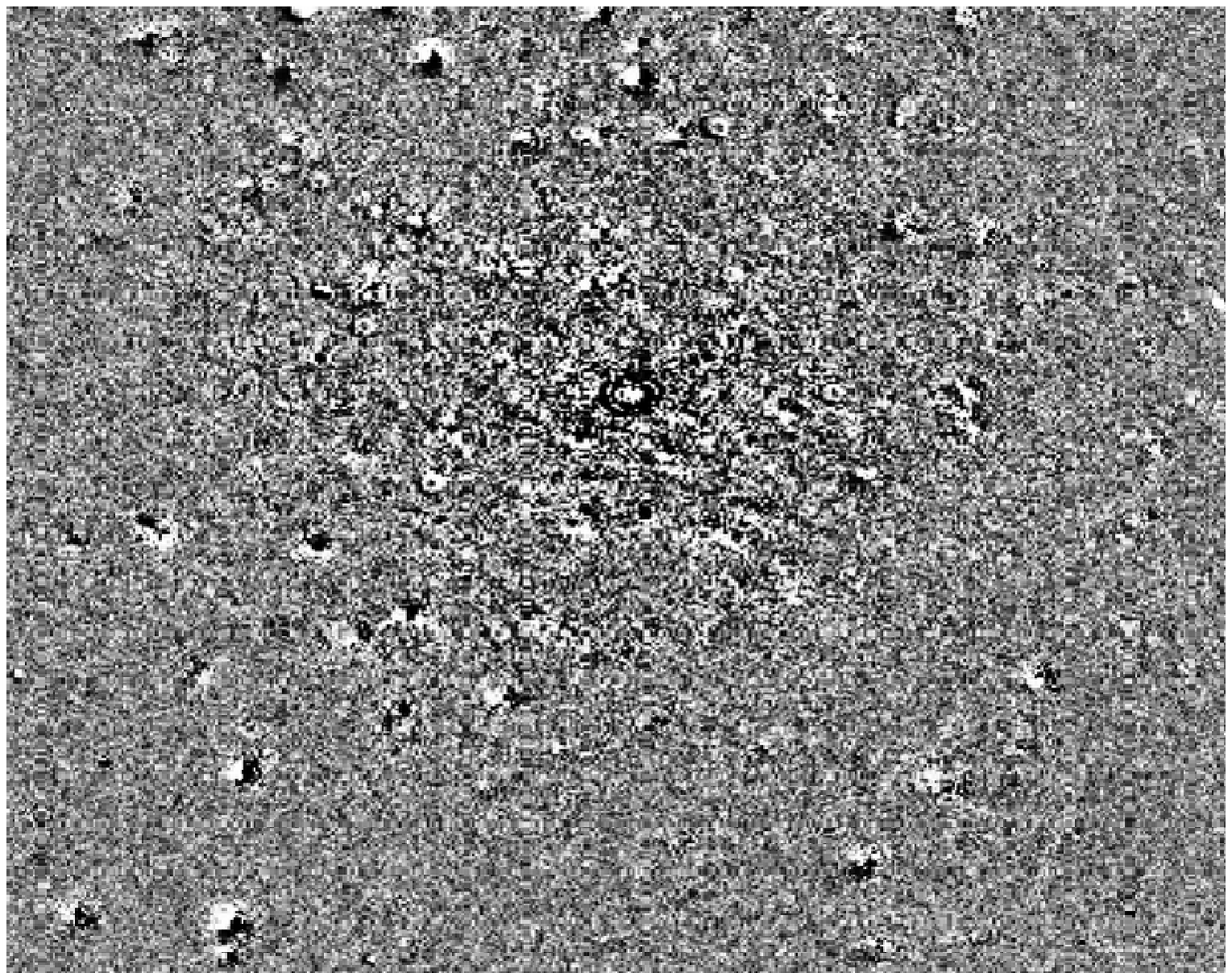}
\caption{{\it Top}: $2 \times 2 $ arcmin central region of the globular cluster 
M15 in V-band (Image-1); {\it Bottom}: The difference between Image-1 and 
Image-2, made with the ISIS package.}
\label{fig:difference}
\end{center}
\end{figure}

\begin{figure}
\begin{center}
\includegraphics[scale=0.25]{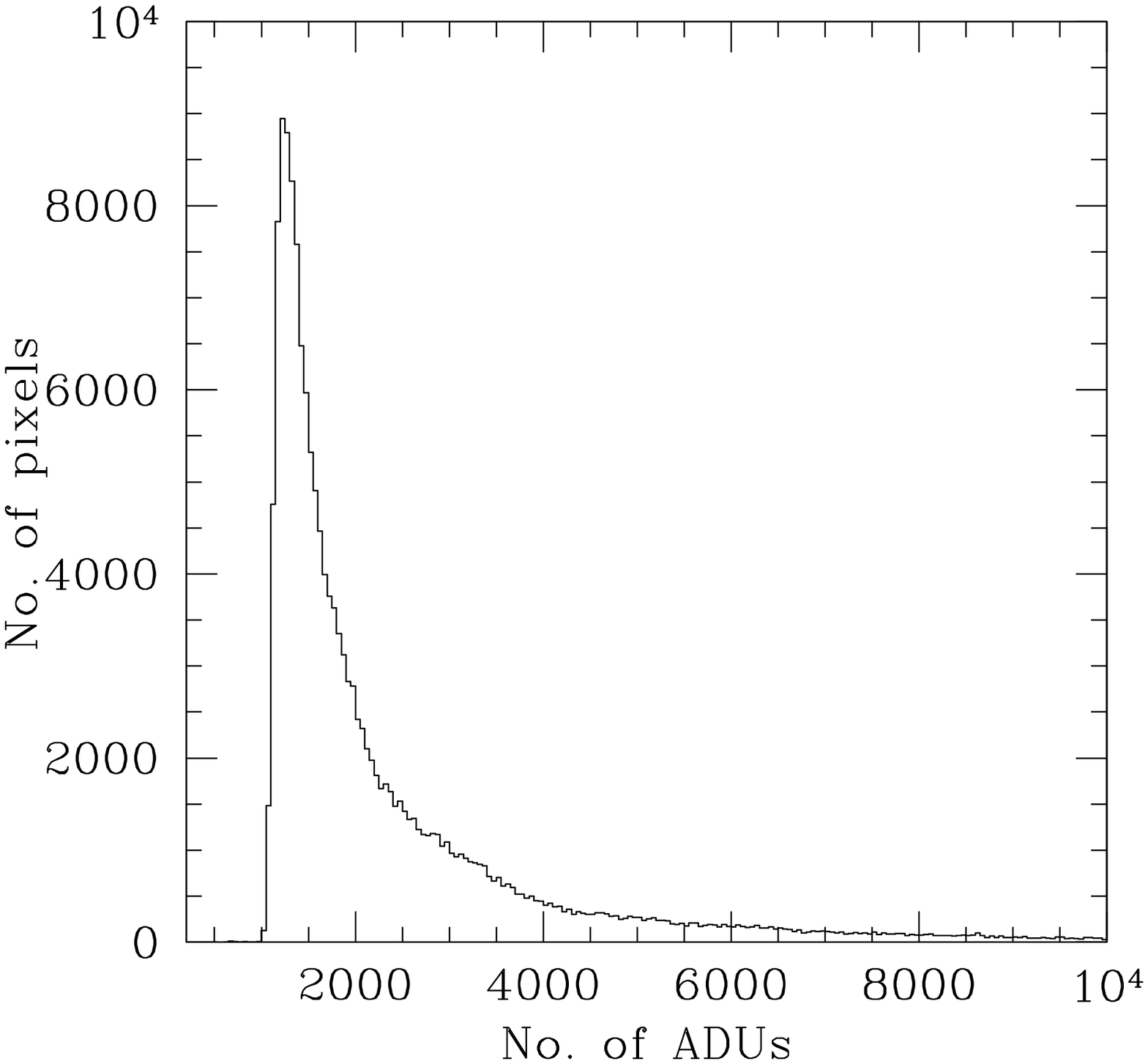}
\includegraphics[scale=0.25]{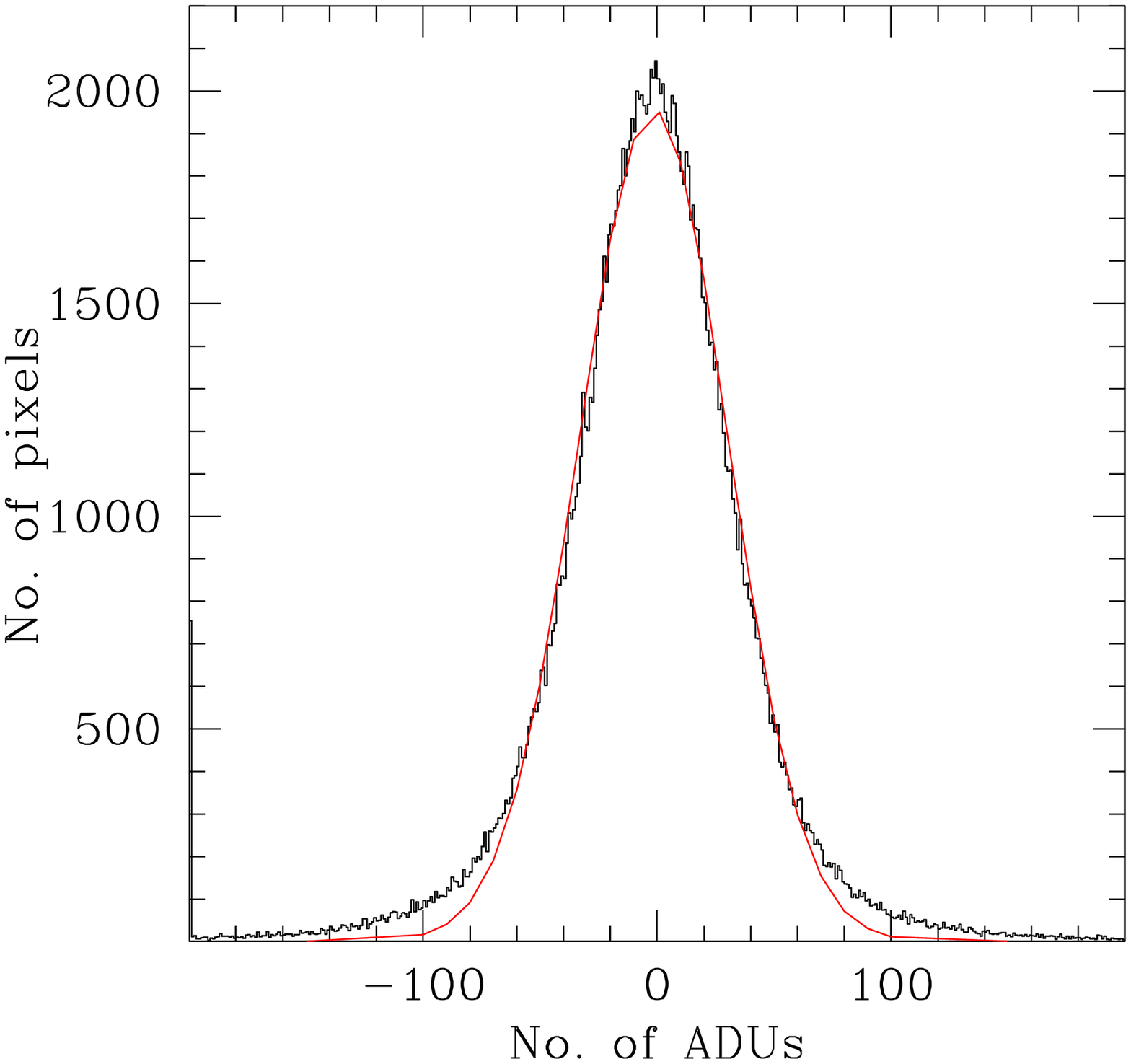}
\caption{{\it Top}: A histogram of the pixel values due to both resolved and unresolved stars
in Image-1; {\it Bottom}:  A histogram of the pixel values in the difference image 
(Fig.~\ref{fig:difference} {\it Bottom}).The solid curve is the Gaussian fit to the distribution.}
\label{fig:histogram}
\end{center}
\end{figure}

In Fig.~\ref{fig:difference} ({\it Top}) we show the central $2\times2$ arcmin 
square region of M15, observed in $V$-band for 50 seconds (Image-1). This 
exposure time was decided to avoid saturation of the core of the cluster in the images. 
In Fig.~\ref{fig:difference}, {\it Bottom} 
we show a difference image between Image-1 and another image taken later 
in the same night (Image-2). The subtraction was done using the ISIS Differential Image 
Analysis package \citep{Alard98}. If a microlensing event of detectable amplitude occurs 
between these two exposure, we can see it appearing as a stellar image in the difference 
image (Fig.~\ref{fig:difference}, {\it Bottom}. This is irrespective of whether the lensed star is 
resolved on not in the original image. However, this residual image may be (and is, indeed) 
affected by the imperfect substraction of bright stars. 
The histogram of the pixel values due to both resolved and unresolved stars 
is shown in the top panel of Fig.~\ref{fig:histogram}.  
In the bottom panel of Fig.~\ref{fig:histogram}, is shown the histogram 
of residual pixel values in this difference image. This distribution is 
nearly gaussian. The standard deviation of this residual pixel value distribution is 
75 ADUs (1 ADU = 1.22 electrons).

\begin{figure}
\begin{center}
\includegraphics[scale=0.3]{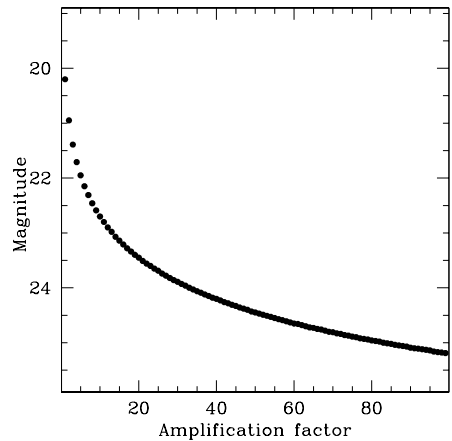}
\caption{{\it HCT} limiting magnitude for detection of stars in 
M15 undergoing lensing event versus lensing amplification.}
\label{fig:amplification}
\end{center}
\end{figure}

The total counts in any star-like image in this difference image are contained 
within a radius of at least 5 pixels (which is around 1.5 arcsecs for {\it HCT}). So the
contribution of noise pixels in this region is 
$\sim 75\times \sqrt{3.14\times 5\times 5} \approx 665$ ADUs. To attain a S/N of $3$, 
the total counts within the stellar PSF have to be $\sim 1995$ ADUs. This  would correspond 
to a limiting magnitude in V-band of 20.2, calculated using the following equation, 
\begin{equation}
m = -2.5 \log{(\mbox{count})} + 2.5 \log{(\mbox{Exp.time)}} + \mbox{Zero-point}\,,
\label{eq:ZP}
\end{equation}
where the zero-point for {\it HCT} in $V$-band is $24.25$ mag ({\bf D. K. Sahu, private 
communication}). Thus, in this observational setup, 
we shall be able to detect a stellar source in the difference image to a magnitude limit of 
$m\approx 20.2$ in V-band. In Fig.~\ref{fig:amplification} we show the HCT limiting magnitude 
for detection of stars in M15 undergoing microlensing for various amplication factors. We would 
like to mention that co-adding several 50 seconds exposures during one night reduces the noise, 
and thereby increases our detection limits.  

It is interesting to mention that, inevitable in the case of a GC microlensing effects of
blending (the contribution to the observed ML light curve from other unassociated sources)
could be used to learn more about the lensing event than would be possible if there
were no blending. The rate of lensing can actually increase in the crowded fields 
\cite{blending95}, because there are many possible lensing events associated 
with the low-mass stars in the resolution cone of the telescope, instead of just one
event associated with lensing of a single isolated star. For example, observing a 
region at $\mu_V=19^{\rm mag} /1\, {\rm arcsec}^2$ increases event probability 
by more than a factor of ten compared to observing in a 14.5 mag region \citep{Colley}. 
Blending plays a 
significant role in making the lensing of lowest mass stars either more or less observable. 
Blending also reduces the duration of events as well, thus it would reduce somewhat the
rather long baseline of observations necessary for central BH cluster microlensing (Table~1).

\subsection{Identifying genuine ML events}

One major difficulty in identifying genuine lensing events is contamination 
by variable stars ({\bf periodic} variables, cataclysmic 
variables (CVs) such as classical novae and dwarf novae (DN)). {\bf Globular
clusters contain very few CVs and especially erupting CVs, which
may mimic microlensing events (Della Valle \& Livio 1996).
The total number as of 2007 was 12 DN in 7 galactic globular
clusters. 
The light curves of variable stars are much different
from symmetric and achromatic Paczynski (1986) curves shown by lensing 
events. Most stellar variables are normally bluer
when they are at maxmimum flux (Sterken \& Jaschek 1996). Thus by
carrying out two colour photometry in V and I bands, and by fitting
the observed light curves to a Paczynski curve,  variable stellar sources
that may cause contamination can be removed.}
The estimated timescales
of GC ML events are of the order of a year, whereas, most of the. 
variable stars in GCs have short periods ($<$ 100 days).
{\bf Therefore, by analysing the observed candidate light curves for the abovementioned
properties, contaminating stellar
variables can be efficiently removed from genuine ML events}.

Self-lensing events may present a challenge due to its higher event
rate (Sec. 2.3). 
{\bf However, as far as
competition between IMBH and self-lensing is concerned, the probabilities are
very different.
The probability of self-lensing does not depend as drastically
on the radial distance as the IMBH lensing. If we consider a tube going through the GC
centre and having, for example, a radius = core radius (which is much more
than the radius of an IMBH influence), the ratio of self-lensing probabilities would
roughly go
as $\tau_{core}/\tau_{allcluster} \propto \frac{1}{2} r_{core}/r_{allcluster}$,
which shows immediately how negligible it would be in the centre.  
Stars behind the cluster centre
cannot contribute to self-lensing optical depth, any self-lensing event there
would only contribute to the photometric noise. Also, contrary to the tentative
conclusion on the choice of GC to look for IMBH lensing, the less centrally
concentrated, loose and diffuse clusters are more favoured for self-lensing
search, and such clusters are excluded from our current proposal.}
In adddition, self-lensing time scales are of the order of days (see Section 2.3), compared to 
the year-long time scales of 
GC ML events (see bottom panel of Fig.~\ref{fig:timeeveprob}). 

GC stars can also be affected by microlensing due to compact matter in the Milky Way 
not associated to a globular cluster \citep{{gyukholder},{Rhoads-Malhotra}}. However, 
in such cases, the lens transverse velocity is set to that of the Galactic disk rotation, 
$v \approx 220$ km/sec, which would give the typical timescale of event 
$\sim 20-30$ days. Such events too will occupy a 
separate region 
on the amplitude-duration diagram and again can be easily distinguised
from long duration ML events and eliminated.

\section{Conclusion}

Determining whether globular clusters contain IMBHs is a key problem in 
astronomy. 
They  may contribute as much as
$\O \approx 0.02$ to the cosmic baryon budget. Their cosmic mass density
could exceed that of supermassive BHs ($\O \approx 10^{-5.7}$) and the
observations do not even rule out that they may account for all the
baryonic dark matter in the Universe ($\O \approx 10^{-1.7}$) \cite{Marel2003}.
They also may have profound influence on the evolution
and survival of globular clusters.
In this paper, we have outlined the novel technique for using 
microlensing to detect IMBH in the centres of GCs. We suggest that dense
and hight luminosity GCs are better suited to search for IMBH than diffuse
and low luminosity GCs. The OH, BD and SC clusters are the most likely ones.

Our suggested monitoring programme of observing $\sim$100 clusters over a 
10 year baseline with a time resolution of 1 observation per month 
 will permit detection of one ML event. Such 
a monitoring programme is feasible on any dedicated 2m class telescope
involved solely in monitoring astronomical sources.
All the potential
contaminants to this lensing signal (Section 3.4), 
can be easily identified and eliminated from the true ML events.

\section{Acknowledgments}
We thank the anonymous referee for his/her helpful comments.
M.~S. would like to thank Sohrab Rahvar with whom this project has started. 
Authors would also like to thank the Indian Institute of Astrophysics and especially our 
Observatories' TACs for providing us with the opportunity to start the project. 


\newpage 

\begin{table*}
\small
\begin{minipage}{130mm}
\caption{Globular cluster candidates for the central black hole;
data and classifications}
\begin{tabular}{|c|cccccccccc|}
\hline \hline 
Cluster & Other & CC$^{\clubsuit}$ & $R_{\odot}$ &
$\s$ & Core radius & $\log{\rho_0}$ & $N_{\rm event}$ & $\la t_{\rm E}\ra$ & Class & Refs \\
name &  &  & $\rm Kpc$ & $\rm km/s$ & $\t_c\,(r_c$, pc) & $\msun\,{\rm pc}^{-3}$ & $10^{-6}$ &days & &  \\
\hline
\hline
NGC104  &47 Tuc& c? & 4.5&11.6&$23.1''$(0.6)&5.0&9.52  &300   &BD& 3,{\bf 4} \\
NGC362&      &c?  & 8.6&6.4 &$0.19'$(0.52)&5.22 &9.39  &507   &YH&{\bf 4},8,9,10\\
NGC1851 &   &  &12.1 & 11.3 &$'$(0.25) &5.7  & 8.0  &  209 &OH&8,10,11 \\
NGC1904 &M79&c?&13.0 & 5.2 &$0.16'$(0.66) &4.2  & 1.5  &  704 &OH&{\bf 4},8,10 \\
NGC2808 &   &  &9.6  &     &$0.25'$(0.73)     &4.9  & 8.9  &  287 &OH  &8,10,11 \\
NGC5139$^{\dagger}$&$\o$ Cen&  & 5.2 &18.47&$2.4'$(3.6) &7.748 &3072.7 &320  &SC&10,20 \\
NGC5272 &M3 &c?&10.4 & 4.8 &$0.55'$(1.26) &3.66 & 1.52 &1053 &YH&1,5,8,10,17 \\
NGC5286 &   &  &11.3 & 8.6 &$0.29'$(0.95) &4.3  & 3.8  & 510 &OH&{\bf 2},5,8,9 \\
NGC5694 &   &  &34.7 & 6.1 &$0.06'$(0.6)  &4.3  & 1.51 & 572 &OH&{\bf 2},5,8,9\\
NGC5824$^{\spadesuit}$ &   &c?&32.0 &11.1 &$0.05'$(0.20) &5.3  & 1.68&182 &OH&{\bf 2},5,8,9 \\
NGC5904 &M5 &  & 7.5 & 6.5 &$0.42'$(0.89) &4.0  & 1.66 & 654 &OH&1,5,8  \\
NGC5946 &   & c&10.6 & 4.0 &$0.08'$(0.25) &4.8  & 0.83 & 563 &OH&{\bf 4},5,8\\
NGC6093 &M80&  &10.0 &14.5 &$0.15'$(0.44) &5.4  & 3.04 & 206 &OH&{\bf 2},5,8,9  \\
NGC6205 &M13&  & 7.7 & 7.1 &$0.78'$(1.75) &3.4  & 1.6  & 839 &OH& 1,8 \\
NGC6256 &Ter12&c & 8.4 & 6.6 &$0.02'$(0.05)&6.6 & 2.1  & 153 &BD&{\bf 4},8,11 \\
NGC6266 &M62&c?& 6.9 &15.4 &$0.18'$(0.36) &5.7  & 13.6 & 176 &OH&1,{\bf 2},5,8\\
NGC6284 &   & c&15.3 & 6.8 &$0.07'$(0.312)&5.2  & 3.24 & 370 &OH&{\bf 4},5,8,9  \\
NGC6293 &   & c& 8.8 & 8.6 &$0.05'$(0.128)&6.3  & 6.9  & 187 &OH&{\bf 4},5,8,9 \\
NGC6325 &   & c& 8.0 &6.4  &$0.03'$(0.07) &6.7  & 5.16 & 186 &OH&{\bf 4},5,8,9 \\
NGC6333 & M9& c& 7.9 &7.59 &$0.58'$(0.91) &4.087& 2.13 & 566 &OH&{\bf 4,}9,10\\
NGC6342 &   & c& 8.6 &5.2  &$0.05'$(0.125)&5.4  & 0.82 & 306 &BD&1,{\bf 4},5,8\\
NGC6355 &   & c& 9.5 &9.02 &$0.05'$(0.14) &4.429& 0.11 & 187 &OH&{\bf 4},9,10\\
Ter 2    &HP3& c& 8.7 & 3.2 &$0.03'$(0.31) &4.261& 0.4  & 784 &BD&{\bf 4},9,10\\
HP 1     &   & c& 14.1&6.35 &$0.03'$(0.58) &4.329& 1.51 & 540 &OH&{\bf 4},9,10\\
Ter 1    &HP2& c& 5.6 & 2.04&$0.04'$(0.3)  &3.891& 0.15 & 1209&YH&{\bf 4},9,10\\
NGC6380 &Ton1&c?&10.7& 6.27&$0.34'$(1.05) &4.64&10.2&736&BD&{\bf 4},9,10\\
NGC6388 &   &  &10.0 &18.9 &$0.12'$(0.35) &5.7  & 73.5 & 337 &BD&{\bf 2},8,9  \\
NGC6397$^{\spadesuit}$ &   &c & 2.4 & 4.5 &$0.05'$(0.2)&5.68 &4.02 & 448 &OH&1,{\bf 2},8,9,14\\
Ter 5    &   &  &10.3 &11.76&$0.18'$(0.54) &6.38 &146.9 &282  &BD&1,10\\
NGC6440 &   &  & 8.4 &9.0  &$'$(0.36)&5.63&11.6  & 300 &BD &10,16,17 \\
NGC6441 &   &  & 11.7&10.0 &$'$(0.5) &5.57& 19.5  & 319 &BD&10,16,17\\
NGC6453 &   & c& 9.6 &6.88 &$0.07'$(0.2)  &4.504&0.27  & 293 &OH&{\bf 4},9,10\\
Ter 6    &HP5& c& 9.5 &7.11 &$0.05'$(0.31) &4.977& 1.9  & 353 &BD&{\bf 4},9,10\\
NGC6522 &   & c&7.8  &7.3  &$0.05'$(0.11) &6.1  & 3.2  & 205 &OH&1,{\bf 4},5,8\\
NGC6540$^{\heartsuit}$
&Djorg3&c?&3.7& 6.0 &$0.03'$(0.03) &5.24&0.03&130&OH&1,{\bf 2},8\\
NGC6541$^{\spadesuit}$ &   &c?&7.0  & 8.2 &$0.3'$(0.13)&5.5  & 1.2 & 198 &OH&1,{\bf 2},8\\
NGC6544 &   &c?&2.7  &9.07 &$0.05'$(0.04) &5.73 & 0.2  & 100 &OH&1,{\bf 4},9,10\\
NGC6558 &   & c&7.4  & 3.5 &$0.03'$(0.07) &5.6  & 0.41 & 341 &OH&{\bf 4},5,8,9\\
NGC6626 &M28&  & 5.6 &8.23 &$0.24'$(0.417) &4.9  & 2.9 & 354 & OH&8,10,11 \\
NGC6642 &   &c?& 8.4 &3.86 &$0.1'$(0.244) &5.244& 2.2  & 577 &YH&{\bf 4},9,10\\
NGC6656 &M22&  & 3.2 &16.99&$1.42'$(1.32) &4.0  & 3.67 & 305 &OH &10,11,17,19  \\
NGC6681 &M70& c&9.2  &10.0&$0.03'$(0.08)  &6.5  & 4.3  & 128 &OH&1,{\bf 4},8 \\
NGC6715$^{\spadesuit \ddagger}$&M54&  &26.3 &20.2  &$0.11'$(0.9)&6.3  &339.4& 212 &SC&{\bf 2},21\\
NGC6717&Pal9&c?& 7.1 & 3.72&$0.08'$(0.13) &5.134& 0.5  & 437 &OH&{\bf 4},9,10\\
NGC6752 &   & c&4.0  &12.4 &$0.17'$(0.2)  &5.2  & 1.33 & 163 &OH&1,{\bf 4},5,8 \\
NGC7078 &M15& c&10.3 & 14.1&$0.07'$(0.2)  &6.87 & 143& 217 &YH&1,{\bf 4},7\\
NGC7099 &M30& c& 8.0 & 5.8 &$0.06'$(0.14) &5.9  & 3.3  & 291 &OH&1,{\bf 4},8\\
G1       &   &c?& 770 & 25.1&$0.21''$(0.78)&5.67 &1080.9& 673 &SC&8,12,15\\
\hline
\end{tabular}

The data in this table is a compilation from various published datasets. 
Where it was possible, we chose the latest available references. 

\emph{Footnotes to the Table~1:} \\
$\clubsuit$  CC: c=post-core-collapse morphology; c?=possible p.c.c. \\
$\heartsuit$  The central density $\rho_0$ was calculated using the formula (Ref.~13) 
$$\rho_0=\fr{3.44\times 10^{10}}{P r_c}10^{-0.4\mu_{\rm V}(0)}
\left(\fr{M}{L_{\rm V}}\right)\,\msun\, {\rm pc}^{-3}\,,$$
with $P\approx 2$ (Ref.~13) and $M/L_{\rm V}=1.1\pm 0.6$ (Ref.~18). The
other values for this formula were taken from Ref.~10.\\
$\spadesuit$ Not likely to host a black hole, according to Ref.~2,
belongs, however, to the CC set. \\
$\dagger$ Instead of $r_c$ radius of BH influence was used for calculations, see Sec.~2.4 \\
$\ddagger$ The latest data is taken from Ibata et al. 2009
\end{minipage}
\end{table*}
\noindent
{\bf REFERENCES TO THE TABLE~1}

\vskip 0.1in
\noindent
1. P.~C.~Freire et al., 2005, ASP Conf. Ser. 328, eds.~F.~A.~Rasio \& I.~H.~Stairs\\
{\bf 2}. Baumgardt H., Makino J., \& Hut P., 2005, \apj, 620, 238.\\
3. P.~C.~Freire et al, 2001, \mnras, 326, 901.\\
{\bf 4}. Trager S.~C., King I. \& Djorgovski S., 1995, \aj, 109, 218.\\
5. P.~Dubath, G.~Meylan and M.~Mayor, 1997, A\&A, 324, 505.\\
6. G.~Meylan and M.~Mayor, 1986, A\&A, 166, 122.\\
7. De Paolis F., Ingrosso G., \& Jetzer Ph., 1996,
\apj, 470, 493.\\
8. C.~Pryor and G.~Meylan, 1993, in {\it Structure and Dynamics of Globular
Clusters}, eds. S.~Djorgovski, G.~Meylan, ASP Conf. Series 50, p. 357.\\
9. Harris Catalogue (Harris, W.E., 1996, AJ, 112, 1487); \\(http://physwww.mcmaster.ca/
\%7Eharris/mwgc.dat) \\
10. Webbink R.~F., 1985, Proc. IAU Symp. 113,
{\it Dynamics of Star Clusters}, eds. Goodman J. \& Hut P., (Kluwer, Dordrecht) p. 541.\\
11. S. Djorgovski, 1993, in {\it Structure and Dynamics of Globular Clusters},
    eds. Djorgovski S, \& Meylan G. ASP Conf. Series, 50, p. 373
12. J.~Ma, et al. 2007, \mnras, 376:1621\\
13. S.~S.~Larsen, 2001, AJ, 122, 1782 \\
14. E.~Dalessandro, B.~Lanzoni, F.~R.~Ferraro, R.~T.~Rood, A.~Milone, G.~Piotto
\& E.~Valenti, 2007, ArXiv e-prints, 712, arXiv:0712.4272.\\
15. G.~Meylan, et al. 2001, \aj, 122:830.\\
16. L.~Origlia, E.~Valenti \& R.~M.~Rich, 2008, \mnras, 388, 1419. \\
17. D.~E.~McLaughlin \& R.~P.~van der Marel, 2005, ApJS, 161:304.\\
18. P.~C\^{o}t\'{e}, 1999, \aj, 118, 406.\\
19. C.~Ding, C.~Li \& W.~Jia-Ji, Chin. Phys. Lett.,2004, 21(8).\\
20. Anderson, J., \& van der Marel, R.~P.\ 2009, arXiv:0905.0627;
Noyola, E., Gebhardt, K., \& Bergmann, M.\ 2008, \apj, 676, 1008.\\
21. Ibata, R., et al.\ 2009, \apjl, 699, L169

\end{document}